\documentclass[10pt]{article}
\usepackage[centertags]{amsmath}
\newcommand{\be}{\begin{equation}}
\newcommand{\ee}{\end{equation}}
\newcommand{\bea}{\begin{eqnarray}}
\newcommand{\nn}{\nonumber}
\newcommand{\eea}{\end{eqnarray}}
\newcommand{\ti}{\widetilde}
\usepackage[centertags]{amsmath}
\usepackage{amsfonts}
\usepackage[]{epsfig}
\begin{document}
%===================================================================
\title{Generalized messenger sector for gauge mediation of supersymmetry breaking and the soft spectrum}
\author{\normalsize  Diego Marqu\'es
\\
{\small\it Departamento de F\'{\i}sica, Universidad Nacional de
La Plata, C.C. 67, 1900 La Plata, Argentina}
\\
{\small\it Associated with
CONICET}
}
\maketitle
%===================================================================
\begin{abstract}
We consider a generic renormalizable and gauge invariant messenger sector and derive the sparticle mass spectrum using the formalism introduced for General Gauge Mediation. Our results recover many expressions found in the literature in various limits. Constraining the messenger sector with a global symmetry under which the spurion field is charged, we analyze Extraordinary Gauge Mediation beyond the small SUSY breaking limit. Finally, we include D-term contributions and compute their corrections to the soft masses. This leads to a perturbative framework allowing to explore models capable of fully covering the parameter space of General Gauge Mediation to the Supersymmetric Standard Model.
\end{abstract}

\section{Introduction}

Supersymmetry (SUSY) solves the hierarchy problem of the Standard
Model making the Higgs mass sensitive to the  scale of soft masses
$m_{soft}$ instead of the Planck scale. Current naturalness criteria
impose a bound for this scale of  new physics $m_{soft} \leq 1\
\rm{TeV}$ and this makes SUSY one of the most appealing candidates to
be found at the LHC. There are different ways to generate soft
terms, being Gauge Mediated Supersymmetry Breaking (GMSB)
\cite{Dine:1981za}-\cite{Dine:1995ag} one of the most studied (see
\cite{Giudice:1998bp}-\cite{Luty:2005sn} for pedagogical reviews).
Among the many advantages of considering this mechanism we can
mention the automatic suppression of the SUSY flavor problem, the
possibility to solve the $\mu$ and SUSY CP  problems  and the fact that it allows the
unification of coupling constants.
After meta-stable dynamical supersymmetry breaking was found to be a
generic phenomenon in ${\cal N} = 1$ SUSY theories
\cite{Intriligator:2006dd}, gauge mediation received renewed
interest. The reason for this is that metastability largely increases the possibilities for model-building in the hidden sector.

As explained in \cite{Meade:2008wd}, general GMSB is not
as predictive as expected. Indeed, predictions of GMSB strongly rely on
 specific hidden-sectors, and are then very model dependent. However, some characteristic features of GMSB are guaranteed, sufficient to distinguish at the LHC
gauge mediation from other popular mediation schemes,
like flavor blindness, gravitino LSP and some
sum rules for sfermions. Other characteristic features that
were supposed to belong to GMSB only arise in the limit of small
SUSY breaking $F \ll X^2$, where $X$ is the typical mass scale of
the hidden sector and $F$ is the strength of the SUSY breaking \cite{Distler:2008bt}.
The general framework of \cite{Meade:2008wd} was also considered in \cite{Ooguri:2008ez}-\cite{Carpenter:2008he} where many different aspects of general GMSB were analyzed.

Specific models capable of covering the
complete spectrum of soft masses, thus leading to general phenomenologies \cite{Carpenter:2008he}, are still lacking. There has been important progress in this direction
\cite{Carpenter:2008wi}-\cite{Buican:2008ws} but a lot remains to be
done. A proof of the existence of models that cover
the whole parameter space of the Supersymmetric Standard Model (SSM)
GMSB was given in \cite{Buican:2008ws}. The main goal of our paper is to provide the mass formulas for gauginos and sfermions that arise in models of that kind.

Let us briefly describe the setup and the results that we obtain in this paper.
We consider a messenger sector defined by a generic renormalizable and gauge invariant mass term for
$N$ messengers
$\phi_i$, $\ti \phi_i$, that may belong to different representations
of different gauge groups
\be W = {\cal M}(X)_{ij}\ \phi_i \ti \phi_j =  (m + X \lambda)_{ij}\
\phi_i \ti \phi_j \label{superpotential}\ee
Here $m$ and $\lambda$ are generic matrices, and $X$ is a spurion field\footnote{There is no loss of
generality when considering a single spurion field
\cite{Cheung:2007es}. If there were more, a unitary transformation
can always be performed such that only one of them acquires an
F-component VEV. The lowest component of the remaining fields can be
absorbed in the matrix $m$ in (\ref{superpotential}).} that acquires a SUSY breaking vacuum expectation
value (VEV)
\be X = X  + \theta^2 F \label{XVEV} \ee
through some unspecified dynamics in the hidden sector that is
irrelevant to our purposes. This tree level coupling to the
messengers provide them with a non-supersymmetric mass; then
integrating out these heavy modes leads to the soft terms. Since
there is no renormalizable tree level coupling with the particles of
the SSM, soft terms arise through loops of messenger and vector
fields. This approach is strongly motivated from the fact that
F-term breaking  models (i.e. generalized O'Raifeartaigh models
\cite{O'Raifeartaigh:1975pr}) arise as low scale limits of
dynamical SUSY breaking theories
\cite{Intriligator:2006dd},\cite{Dine:2006gm}-\cite{Dine:2007dz}.

Applying the formalism of general GMSB \cite{Meade:2008wd} we derive
the sparticle soft spectrum arising after integration of our
generalized messengers. The formulas we obtain generalize the results
in \cite{Dimopoulos:1996gy}-\cite{Giudice:1997ni}, since they include all mixing effects due to multiple messenger scales, and hold for arbitrary amount of SUSY breaking. We then constraint
the messenger sector with a global (R or non-R) symmetry and
 show how Extraordinary Gauge Mediation \cite{Cheung:2007es} behaves beyond the limit of small
SUSY breaking. We find bounds on the deviations from this limit, analogous to those in
\cite{Martin:1996zb}. The possibility of
 ``diagonal type'' splitting  between
fermion and boson messenger masses resulting from D-terms is also considered. We show that in this case some non positive
definite contributions arise in the soft spectrum, that highly
modify the relations among sparticles and are crucial for
constructing models that span the full parameter space of general
GMSB.

The outline of the paper is as follows. In Section \ref{Formulas} we
present the general expressions for the sparticle masses, and show
some limits previously considered in the literature. In Section
\ref{EOGM} we constraint the messenger sector with a non-trivial
global symmetry and analyze Extraordinary Gauge Mediation beyond the
small SUSY breaking limit. Section \ref{Dterm} is devoted to analyze D-term
contributions and to show how the soft spectrum gets affected.
Finally, Section \ref{Conclu} concludes, and in the Appendix we
present a detailed computation of the sparticle masses.

\section{Generalized sparticle mass spectrum}\label{Formulas}

Following the general GMSB formalism \cite{Meade:2008wd}, soft masses for gauginos and sfermions are derived in the Appendix. These formulas are exact at one loop for gauginos and at two loops for sfermions. Since the messenger sector is the most general of its kind, our expressions generalize many of those found in the literature. In this section we show that the obtained masses recover known formulas in various different limits (sparticle spectroscopy beyond the minimal framework was also studied in \cite{Dimopoulos:1996yq}-\cite{Gorbatov:2008qa}). In Section \ref{Dterm} we will extend these results so as to include effects from D-term SUSY breaking.

Without loss of generality, after a redefinition of the messenger superfields we choose a basis in which $\cal M$ in (\ref{superpotential}) is
diagonal with real eigenvalues $m_k^0$. In this basis, $F \lambda$ is still a generic matrix in flavor space. As we explain in the Appendix, dangerous negative contributions to sfermion masses can arise from D-terms at one-loop, unless we impose a symmetry that constraints the messenger sector in such a way that there exists a basis in which $\cal M$ is diagonal and $F \lambda$ hermitic. In this paper we assume the messenger sector to be constrained in such a way and for default stand in a basis that diagonalize $\cal M$, except in Section \ref{EOGM} where it is more convenient to stand in a different basis.

We define the unitary matrices $U_\pm$ as those that diagonalize ${\cal M}^2 \pm F \lambda$, the hermitic mass-squared matrix for bosonic messengers, namely
\be {\cal M}^2_\pm = U_\pm^\dag ({\cal M}^2 \pm F
\lambda) U_\pm \label{Diagonalizing}\ee
Then, ${\cal M}^2_\pm$ are diagonal matrices with real eigenvalues $(m_k^\pm)^2$. We also define the following mixing matrices
\be A_{kn}^\pm = (U^\dag_\pm)_{kn} (U_\pm)_{nk} \ , \ \ \ \ \ B_{kn}^\pm = (U^\dag_\pm U_\mp)_{kn} (U_\mp^\dag U_\pm)_{nk}
 \ee

\subsection{Gaugino masses}
In the Appendix a detailed computation for the gaugino masses $M_r$ at the messenger scale is presented, with the result
\be M_{r} = \frac{\alpha_r}{4 \pi}
\Lambda_{G} \ , \ \ \ \ \Lambda_{G} = 2 \sum_{k,n = 1}^N \sum_\pm \pm\ d_{kn}\ A_{kn}^\pm\ m^0_n  \frac{(m_k^\pm)^2 \log ((m_k^\pm)^2/(m_n^0)^2)}{(m_k^\pm)^2 - (m_n^0)^2} \label{LambdaG}\ee
Here $d$ is the Dynkin coefficient for the messengers, $d\delta^{ab} = {\rm Tr} [T^a T^b]$, in a normalization where $d = 1/2$ for ${\bf N_c} + {\bf \bar N_c}$ bi-fundamentals of $SU(N_c)$ \footnote{The indices $k$ and $n$ run over messengers. The Dynkin index $d_{kn}$ is nonzero only when $k$ and $n$ label fields that are in the same representation.}. For short, we omit the label $r$ of the gauge group in the gaugino and sfermion scales, $\Lambda_{G r}$ and $\Lambda_{S r}$ respectively.

In the Minimal Gauge Mediation (MGM) limit \cite{Martin:1996zb} which we obtain by setting $m = 0$ in equation (\ref{superpotential}), this reduces to
\be \Lambda_G = \frac{F}{X}  \sum_{k=1}^N 2 d_{kk}\  g (x_k) \label{LGMGM} \ee
\be g(x) = \frac{(1 +
x) \log(1 + x)}{x^2} + (x \to -x) \label{ge}\ee
where we have defined $x_{k} = \frac{F}{\lambda_k X^2}$, $\lambda_k$ being
 the eigenvalues of $\lambda$. To lowest order in $F/X^2$ this MGM expression was obtained through the Wave-function Renormalization Technique (WRT) \cite{Giudice:1997ni}.

The lowest order in the $F/X^2$ expansion of (\ref{LambdaG}) coincides with the result of \cite{Dimopoulos:1996ig} for the case of $SU(N_c)$ bi-fundamentals
\be
\Lambda_G = \sum_{k = 1}^N  \frac{F\lambda_{kk}}{m^0_k} = \partial_X \log \det {\cal M} \label{GparaGiuYCheung}
\ee
In the last equality of equation (\ref{GparaGiuYCheung}) we call $\cal M$ to the original matrix in (\ref{superpotential}) before diagonalization, and use the fact that in the original basis the identity $\partial_X {\cal M} = \lambda$ always holds. This allows us to write $\Lambda_G$ as in \cite{Cheung:2007es} where this expression was derived using a generalization of the WRT.

The generalized expression (\ref{LambdaG}) can be approximated by
\be
\Lambda_G =  \sum_{k = 1}^N 2 d_{kk}\ \frac{F \lambda_{k}}{m_k^0} \  g\left(\frac{ F \lambda_{k}}{(m_k^0)^2}\right)\label{ExpansionFGauginos}
\ee
but this gets corrected by multi-messenger mixing effects arising at order ${\cal O}(F / X^2)$ in the expansion of $g(x)$, that vanish in the limit ${\cal M} \to m^0 1_{N\times N}$, i.e. $m_k^0 \approx m^0, \forall k$. In this limit the fermionic messengers become degenerate but the bosons are still arbitrarily split, so SUSY can still be largely broken in which case the messenger scales will lie in a large range.

Let us finally mention that the equation (\ref{LambdaG}) is given at the messenger scale and must be renormalized down to the scale of SSM particles. At the electroweak scale we find that the correction at the leading order ${\cal O}(\alpha)$ only comes from the running of the coupling constant

\be M_{r}(Q) = \frac{\alpha_r(Q)}{4 \pi} \Lambda_{G r}\ee

\subsection{Sfermion masses}

The sfermion masses at the messenger scale can be written as usual
\be m_{\ti f}^2 = 2 \sum_{r =1}^3 C_{\ti f}^r  \ \left(\frac{\alpha_r}{4\pi}\right)^2 \ \Lambda_{S}^2 \label{masafermiones}\ee
where $C_{\ti f}^r$ are the quadratic Casimir of $\ti f$ in the gauge group $r$. The sfermion scales $\Lambda_S^2$ for the model (\ref{superpotential}) have been computed in the Appendix, and read
\be \Lambda_S^2 = 2 \sum_{k,n = 1}^N \sum_\pm d_{kn}   (m_k^\pm)^2
\left[\ A^\pm_{kn} \log \frac{(m_k^\pm)^2}{(m_n^0)^2} - 2\ A^\pm_{kn}\ Li_2
\left(1 - \frac{(m^0_n)^2}{(m_k^\pm)^2}\right) +\ \frac{1}{2}\ B^\pm_{kn}\ Li_2 \left(1 - \frac{(m_n^\mp)^2}{(m_k^\pm)^2}\right)
\right]\label{LambdaS}\ee
$Li_2$ being the dilogarithm function.
Again, by setting $m = 0$ in equation (\ref{superpotential}), the MGM limit is recovered \cite{Dimopoulos:1996gy}, \cite{Martin:1996zb}
\be \Lambda_S^2 =  \left(\frac{F}{X}\right)^2 \sum_{k=1}^N 2
d_{kk} f(x_k)\label{LSMGM}\ee

\be f(x) = \frac{1
+ x}{x^2} \left[ \log(1 + x) - 2  Li_2 \left(\frac{
x}{1 + x}\right)  + \frac{1}{2} Li_2 \left(\frac{2 x}{1
+ x}\right)\right] + (x\to -x)\label{efe}\ee

In the limit of small multi-messenger mixing effects (which arise in this case at order ${\cal O}(1)$ in the $F / X^2$ power expansion of $f(x)$) we obtain from (\ref{LambdaS}) the result
\be \Lambda_S^2 =   \sum_{k = 1}^N 2 d_{kk}\ \frac{F^2 \lambda_{k}^2}{(m_k^0)^2} \ f\left(\frac{F \lambda_{k}}{(m_k^0)^2}\right) \label{ExpansionFSfermions}\ee

In the approximation of small multi-messenger mixing effects, to lowest order in the $F/X^2$ expansion, for the case of $SU(N_c)$ bi-fundamentals, the results of \cite{Cheung:2007es} are recovered from equation (\ref{ExpansionFSfermions}), namely
\be \Lambda_S^2 = \frac{1}{2}F^2 \frac{\partial^2}{\partial X
\partial X^*} \sum_{k=1}^N \log^2 (m^0_k)^2 \label{EOGMS}\ee

Setting ${\cal M} = m^0 1_{N\times N}$ in (\ref{LambdaS}), to lowest order in $F/X^2$ for $SU(5)$ bi-fundamentals, we reproduce the result of \cite{Dvali:1996cu},\cite{Dimopoulos:1996ig}
\be
\Lambda_S^2 = \sum_{i,j = 1}^N \frac{|F \lambda_{ij}|^2}{(m^0)^2}
\ee

As in the case of gaugino masses, sfermion masses are given here at the messenger scale and must be RG evolved down to the soft scale. General expressions for this evolution can be found in \cite{Martin:1993zk}, \cite{Cheung:2007es}, being the leading contribution to order ${\cal O}(\alpha^2)$
\be
\frac{d m_{\ti f}^2}{d \log Q} \sim \sum_{r = 1}^3 8 \frac{\alpha_r^2}{(4 \pi)^2}\ C_{\ti f}^r\ {\rm Str}_k (d^r_{kk} M^2_{r, k}) \ee
where $M$ is the complete messenger mass matrix including bosons and fermions. Notice that due to the Supertrace Theorem, these corrections have no effects above the ultimate messenger threshold.

\section{Constraining the messenger sector}\label{EOGM}

When the messenger superpotential (\ref{superpotential}) is constrained with a global symmetry under which the
spurion field $X$ is charged, these models become those of ExtraOrdinary Gauge Mediation (EOGM) recently analyzed
in \cite{Cheung:2007es}. The addition of this single symmetry leads to different type of models, each of which
have characteristic features that make them highly predictive despite their complexity. Some of the interesting features
that these models present are modified relations between squark and slepton masses, the possibility for small $\mu$
and Higgsino NLPS, effective messenger number less than one, and gauge coupling unification.
In addition to  briefly reviewing EOGM, in this section we address the following two issues. First generalize the definition of EOGM by showing that an R symmetry is indistinguishable from a non-R symmetry in the messenger sector. Then we analyze these models in the large SUSY breaking limit and show how they deviate from the small $F/X^2$ behavior, which is the regime in which they were studied in \cite{Cheung:2007es}.

In this section we do not stand in the basis in which $\cal M$ is diagonal, but in the ``original'' basis in which the global symmetry is evident. We consider the most general messenger mass superpotential consistent with gauge invariance, renormalizability and a global $U(1)$ symmetry which can be R or non-R. The messenger content consists of $N$ messengers in the ${\bf 5} + {\bf \bar 5}$ representation of $SU(5)_{GUT}$. The superpotential has the form %
\be W = {\cal M}(X)_{ij}\ \phi_i \ti \phi_j =  (m + X \lambda)_{ij}\
\phi_i \ti \phi_j\ee
and is invariant under a global symmetry $G$ under which $X$ carries a non-vanishing charge, which we take to be positive $G(X) > 0$ without loss of generality. If $G(W) \neq 0$ we are in the case of an R-symmetry, and otherwise if $G(W) = 0$, the symmetry is non-R. This symmetry implies the following selection rules for $m$ and $\lambda$
\bea
m_{ij} \neq 0 &\Longrightarrow& G(\phi_i) + G(\ti \phi_j) = G(W)\nn\\
\lambda_{ij} \neq 0 &\Longrightarrow& G(\phi_i) + G(\ti\phi_j) = G(W) - G(X)\label{selection}
\eea
The model always has in addition an accidental trivial global $G'$ symmetry under which $G'(X) = 0$, $G'(\phi_i) + G'(\ti \phi_j) = G'(W)$.

After spontaneous breaking of GUT symmetry, the messenger sector can split in doublet and triplet sectors, each containing its own mass matrix ${\cal M}_{2}$ and ${\cal M}_{3}$ respectively
\be W = {\cal M}_{2 ij}\ \phi_i^2 \ti \phi_j^2 + {\cal M}_{3 ij}\ \phi_i^3 \ti \phi_j^3 \label{superpotentialsplit}\ee
$\phi^r, \ti \phi^r$ being  in the ${\bf r} + \bf {\bar r}$ representation of $SU(r)$, $r = 2,3$, the doublet and triplet components of the $\bf 5$-plet messengers $\phi, \ti \phi$. Each sector then interacts separately with weakly and colored coupled matter.

Following a route completely analogous to that in \cite{Cheung:2007es} one can prove that due to the G-symmetry the following identity for the determinant of the messenger mass matrices holds
\be
\det {\cal M}_r = X^{n_r} f(m_r, \lambda_r)\ , \ \ \ \ \ \ \ n_r \equiv \frac{1}{G(X)} \sum_{i = 1}^N (G(W)- G(\phi_i^r) - G(\ti \phi_i^r))\ , \ \ \ \ \ \ \ 0\leq n_r \leq N \label{detM}
\ee
This identity is related to gauge couplings unification. In fact, every scale of the hidden sector alters the running of the couplings, producing the following final shift at the GUT scale
\be
\delta\alpha_r^{-1} = - \frac{N}{2\pi} \log\frac{m_{GUT}}{\bar{\cal M}_r}\ ,\ \ \ \ \ \ \ \ \bar{\cal M}_{2,3} \equiv (\det {\cal M}_{2,3})^{1/N}\ ,\ \ \ \ \ \ \ \ \bar {\cal M}_1 \equiv (\bar{\cal M}_2)^{3/5} (\bar{\cal M}_3)^{2/5}
\ee
Couplings unify when the condition $\bar {\cal M}_2 = \bar {\cal M}_3$ is satisfied, which in turn implies $n_1 = n_2 = n_3$ in (\ref{detM}). This relation $\bar {\cal M}_2 = \bar {\cal M}_3$ (and then unification) can be maintained even in the presence of large doublet/triplet splitting, since the function $f$ in (\ref{detM}) is generally independent of some subset of the parameters.

In \cite{Cheung:2007es}, these models have been classified in three distinct classes depending on the specific details of the matrices $m$ and $\lambda$. Here we define them and mention their characteristic features (in the following items we omit
 the gauge group index $r$)
\begin{itemize}
\item {\bf Type I.} Models with $\det m \neq 0$.

In these models there always exists a basis in which $m$ is diagonal, and fields must come in pairs with G-charges $G(\phi_i) + G(\ti \phi_i) = G(W)$. In such basis we order the fields $\phi_i$ in increasing G-charge, which allows us to rewrite the selection rules (\ref{selection}) as
    \bea
    &&\lambda_{ij} \neq 0 \Longrightarrow G(\phi_i) - G(\phi_j) = - G(X)\ \ (\Rightarrow\ \  j > i)\label{Desde}\\
    && m \ {\rm diagonal}
    \eea
    These rules imply that $\lambda$ is strictly upper diagonal, so in these models the eigenvalues of $\cal M$ are those of $m$ and then $\det {\cal M} = \det m$ and $n = 0$. The messengers are stable in a neighborhood of $X = 0$, but can become tachyonic for large $X$. Examples of Type I models are \cite{Dine:1981gu}-\cite{Dimopoulos:1982gm}, and the more recent ISS model \cite{Intriligator:2006dd} and many of its variations \cite{Kitano:2006wm}-\cite{Abel:2008gv}.

\item {\bf  Type II.} Models with $\det \lambda \neq 0$.

In these models there always exists a basis in which $\lambda$ is diagonal, and fields must come in pairs with G-charges $G(\phi_i) + G(\ti \phi_i) = G(W) - G(X)$. In such basis we order the fields $\phi_i$ in decreasing G-charge, which allows us to rewrite the selection rules (\ref{selection}) as
    \bea
    && m_{ij} \neq 0 \Longrightarrow G(\phi_i) - G(\phi_j) = G(X)\ \ (\Rightarrow\ \  j > i)\\
    && \lambda \ {\rm diagonal}
    \eea
    These rules imply that $m$ is strictly upper diagonal, so in these models the eigenvalues of $\cal M$ are those of $X\lambda$ and then $\det {\cal M} = X^N \det \lambda$ and $n = N$. The messengers are tachyonic in a neighborhood of $X = 0$, but can become stable for large $X$.

\item {\bf Type III.} Models with $\det m = \det \lambda = 0$.

Since from (\ref{detM}) $\det {\cal M}\sim X^n$ is a monomial, the eigenvalues of $\cal M$ are either proportional to $X$ or independent of $X$. Then, analyzing the limits $|X| \to 0,\infty$, one reaches the conclusion that the eigenvalues of $\cal M$ are the $n$ non-vanishing eigenvalues of $X \lambda$ and the $N - n$ non-vanishing eigenvalues of $m$. We can then choose a basis in which the eigenvalues of $X \lambda$ appear in the first $n$ entries of the diagonal,  and the eigenvalues of $m$ in the last $N - n$ entries. From (\ref{selection}) we see that fields come in
  pairs with G-charges $G(\phi_i) + G(\ti \phi_i) = G(W) - G(X)$ for $1 \leq i \leq n$, and $G(\phi_i) + G(\ti
    \phi_i)  = G(W)$ for $n < i \leq N$. Ordering the first $n$ fields $\phi_i$ in decreasing order of G-charge, and the last $N - n$ in increasing order, the selection rules (\ref{selection}) can be written for the following four blocks of $\cal M$
    \bea
    1 \leq i ,j \leq n &:& m_{ij} \neq 0 \Longrightarrow G(\phi_i) - G(\phi_j) = G(X) \ \ (\Rightarrow\ \ j > i )\\
    && \lambda\ {\rm diagonal} \nn\\
    n < i ,j \leq N &:& \lambda_{ij} \neq 0 \Longrightarrow G(\phi_i) - G(\phi_j) = - G(X) \ \ (\Rightarrow\ \ j > i )\\
    && m\ {\rm diagonal} \nn \\
    i \leq n < j &:& m_{ij}\neq 0 \Longrightarrow G(\phi_i) = G(\phi_j)   \\
    && \lambda_{ij}\neq 0 \Longrightarrow G(\phi_i) - G(\phi_j) = - G(X)\nn \\
     j \leq n < i &:& m_{ij}\neq 0 \Longrightarrow G(\phi_i) - G(\phi_j) = G(X) \label{Hasta} \\ && \lambda_{ij}\neq 0 \Longrightarrow G(\phi_i) = G(\phi_j)\nn
    \eea
 Since $\det \lambda = \det m = 0$ these models have $0 < n < N$. The messengers are tachyonic in the limits $|X|  \to 0, \infty$, but can become stable in some intermediate region.
\end{itemize}

Notice that we have been able to write the selection rules (\ref{Desde})-(\ref{Hasta}) in a way completely independent of $G(W)$, so at the level of the messenger mass matrix an R-symmetry is undistinguishable from a non-R symmetry. In other words, a reassignment of G-charges leads from one type of symmetry to the other. Where the nature of the symmetry becomes manifest is in the completion of the messenger sector. When trying to complete this superpotential to a SUSY breaking hidden sector, an R-symmetry leads to simple models of direct gauge mediation \cite{Cheung:2007es}, while a non-R symmetry forbids $X$ to acquire a non-vanishing F-component (\ref{XVEV}) \cite{Nelson:1993nf}. We also learn from the above results that the global symmetry constraining the superpotential implies that the eigenvalues of $\cal M$ are those of $m$ and $X \lambda$ \footnote{There are other non-generic constraints that enforce the eigenvalues of $\cal M$ to be those of $m$ and $X\lambda$ \cite{Marques:2008va}.}, and then $\lambda_k = \partial_X {\cal M}_k$. This fact can be useful when analyzing sparticle masses in these models, in particular gaugino masses.

In the limit of small multi-messenger mixing effects and to leading order in $F/X^2$ The expressions for gauginos and sfermions are
\bea
&&M_{r} = \frac{\alpha_r}{4\pi} \Lambda_{G r} \ , \ \ \ \ \ \ \ \ \ \ \ \ \ \ \ \ \ \ \ \ \ \ \ \ \ \ \ \ \ \ \ \ \ \ \ \Lambda_{G r} = F \partial_X \log \det {\cal M}_r = n_r \frac{F}{X}\label{FormulasMasas1}\\
&& m_{\ti f}^2 = 2 \sum_{r =1}^3 C_{\ti f}^r \ \left(\frac{\alpha_r}{4\pi}\right)^2 \Lambda_{S r}^2 \ , \ \ \ \ \ \ \ \ \ \ \ \ \ \ \ \ \  \Lambda_{S r}^2 = \Lambda_{G r}^2\ \ti N^{-1}_r \label{FormulasMasas2}\eea
where $C_{\ti f}^r$ are the quadratic Casimir of $\ti f$ in the gauge group $r$, and we have defined the effective messenger number
\be \ti N_{r}(X) \equiv \left[\frac{1}{2n_r^2} |X|^2 {\partial^2_{XX^*}} \sum_{k=1}^N \log^2 (m_k^0)_r^2 \right]^{-1} \label{Neff}\ee

As pointed out in \cite{Cheung:2007es}, the gaugino masses are proportional to the index $n_r$ defined in (\ref{detM}), which depends on the G-charges of the doublet and triplet messengers. When the $SU(2)$ and $SU(3)$ multiplets are given the same G-charges, we have $n_1 = n_2 = n_3$ and then unifications of coupling constants and an MGM spectrum for gauginos (including gaugino mass unification) is achieved. Let us suppose that this is the case so as to simplify the analysis of the sfermion spectrum.

The effective messenger number $\ti N$ defined in (\ref{Neff}) does not necessarily coincide with the number of messengers $N$ (as it is the case in MGM), but is a more general function of $X$. In fact, its asymptotic behavior in the limits $|X| \to 0, \infty$ is independent of all the parameters, and is bounded by
\bea
\frac{n_r^2}{n_r^2 - (N - r_{m_r} - 1)(2n_r - N + r_{m_r})}\ \leq &\ti N_r (X\to 0)& \leq\ N - r_{m_r} \nn\\
\frac{n_r^2}{r_{\lambda_r} + (r_{\lambda_r} - n_r)^2}\ \leq &\ti N_r (X\to \infty)& \leq\ \frac{n_r^2}{r_{\lambda_r} + \frac{(r_{\lambda_r} - n_r)^2}{N - r_{\lambda_r}}}  \label{BoundsNeff}
\eea
where $r_{m_r} \equiv {\rm rank}\ m_r$ and $r_{\lambda_r} \equiv {\rm rank}\ \lambda_r$.
Since we have assumed that the $n_r$'s are equal, the splitting among sfermions is controlled (in addition to the couplings $\alpha_r$) by this effective messenger number $\ti N_r$, as can be seen from equations (\ref{FormulasMasas1})-(\ref{FormulasMasas2}). The mass relations amongst sfermions can then be arbitrarily modified with respect to MGM.

When going beyond the small SUSY breaking limit, $F \lambda_k / (m_k^0)^2$ need not be much smaller that $1$ and should not be sufficiently close to $1$, since otherwise some bosonic messengers would be too light. To analyze this case, expressions (\ref{ExpansionFGauginos}) and (\ref{ExpansionFSfermions}) should be used, where the functions $g(x)$ and $f(x)$ were defined in (\ref{ge}), (\ref{efe}).
\begin{figure}\begin{center}
\includegraphics[width=10 cm]{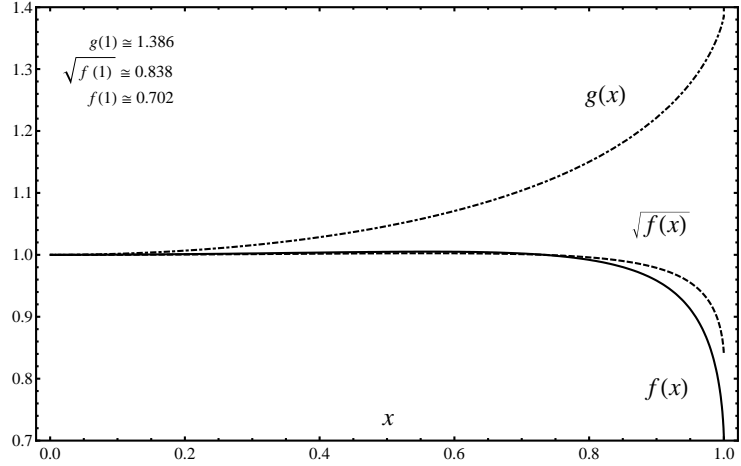}
\end{center}
\caption{\small $g(x)$, $f(x)$ and $\sqrt{f(x)}$ are functions of $F \lambda_k / (m_k^0)^2$ (a parameter related to the amount of SUSY breaking) that weigh the contribution of each messenger to the spectrum.} \label{Valley}
\end{figure}
The expansion of functions $g(x)$ and $f(x)$ give very good accuracy to order ${\cal O}(x^6)$
\be
g(x) = 1 + \frac{x^2}{6} + \frac{x^4}{15}+ \frac{x^6}{28} + \dots \ , \ \ \ \ \   f(x) = 1 + \frac{x^2}{36} - \frac{11}{450} x^4 - \frac{319}{11760} x^6 + \dots
\ee
except when $x$ is very close to $1$, which is a possibility that we have discarded. Interestingly, as can be seen in Figure 1, neither function deviates sufficiently from $1$ so as to alter drastically the spectrum. In fact we automatically obtain the bounds
\be
\frac{F}{X}\ n_r \leq \Lambda_{G r} \leq 1.386\ \frac{F}{X}\ n_r\ , \ \ \ \ \ 0.838\ \frac{n_r}{\sqrt{\ti N_r}}\ \frac{F}{X} \leq \Lambda_{S r} \leq \frac{n_r}{\sqrt{\ti N_r}}\ \frac{F}{X}  \label{bounds}
\ee
which combined yield
\be
\ti N_r \leq \frac{\Lambda_{G r}^2}{\Lambda_{S r}^2}\leq 2.735\ \ti N_r \label{bounds2}
\ee
where $\ti N$ is defined in (\ref{Neff}), as the quotient between the gaugino and scalar fermion scales in the $F \lambda_k \ll (m^0_k)^2$ limit. We can define an effective non-integer $\ti n_r \equiv X \Lambda_{G r} / F$ which due to equation (\ref{bounds}) will be bounded as $n_r\leq \ti n_r\leq 1.386\ n_r$. Since the integers $n_r$ depend on the G-charge assignments uniquely (\ref{detM}), it is possible to achieve arbitrary doublet/triplet splitting together with gaugino mass unification in the limit of small SUSY breaking. The numbers $\ti n_r$ are on the contrary a function of all the parameters. Due to this, gaugino mass unification is lost in the large SUSY breaking limit
\be
M_1 : M_2 : M_3 = \alpha_1 \ti n_1 : \alpha_2 \ti n_2 : \alpha_3 \ti n_3
\ee

The bounds (\ref{bounds})-(\ref{bounds2}) can be further tightened lowering the value of $F \lambda_k / (m^0_k)^2$, and also considering that there are generally some $\lambda_k$'s in EOGM that identically vanish.

We conclude that the spectrum of Extraordinary Gauge Mediation models responds to variations in the SUSY breaking quotient $F/X^2$ in a way similar to that of MGM. Namely, it is quite insensitive to the value of $F / X^2$ except when this value is very close to $1$. These results hold only in the limit of small effects arising from mixing of the multi-messenger scales.

\section{D-term contributions}\label{Dterm}

Realistic models at hand do not span the most general spectrum that can be achieved through gauge mediation \cite{Meade:2008wd}. A program to construct models allowing to cover such space (and therefore allowing to explore general phenomenologies \cite{Carpenter:2008he}) was started in \cite{Carpenter:2008wi}. In that paper the models have the correct number of variables, but do not span the general spectrum. More recently, a proof of the existence of specific models that explore the full parameter space of general GMSB was given in \cite{Buican:2008ws}, where in addition to F-breaking, the possibility for D-breaking was considered. In this section we will provide complete formulas of soft masses for such models.

We start by considering the (SUSY broken) messenger mass matrices of \cite{Buican:2008ws}
\be
M_F = \left(\begin{matrix} 0 & {\cal M}^\dag \\ {\cal M} & 0\end{matrix}\right)\ , \ \ \ \ \ \ \ \ M_B = \left(\begin{matrix} {\cal M}^\dag{\cal M} + \xi & (F \lambda)^\dag \\ F\lambda & {\cal M}{\cal M}^\dag  + \ti \xi \end{matrix}\right) \label{matrizconxi}
\ee
with $\xi, \ti \xi$ hermitic matrices, that can arise for example when messengers are charged under a $U(1)$ gauge group with a non-vanishing Fayet-Illiopulos D-term \cite{Nakayama:2007cf}-\cite{Luo:2008zr}, or also through some strong gauge dynamics \cite{Seiberg:2008qj}. They correspond to ``diagonal type'' terms of the form
\be
V \supset \xi_{ij}\  \phi_i\phi_j^\dag + \ti \xi_{ij}\ \ti \phi_i \ti \phi_j^\dag
\ee
in the potential.

As we have done previously, we stand in a basis in which $\cal M$ is diagonal and real, and impose a messenger parity. Such symmetry constraints $F \lambda$ to be hermitic, and imposes $\xi = \ti \xi$. By demanding that the theory be insensitive to UV physics, we impose the supertrace of the whole messenger mass matrix to be vanishing \cite{Poppitz:1996xw}, \cite{Buican:2008ws}, which in this case accounts for ${\rm Tr} [d\ \xi] = 0$. Under these conditions, exactly all the results (\ref{supercurrent})-(\ref{G4}) of the appendix, and expressions (\ref{LambdaG}), (\ref{masafermiones})-(\ref{LambdaS}) of Section \ref{Formulas} hold, except for the fact that $(m_k^\pm)^2$ are now the eigenvalues of ${\cal M}^2 \pm F \lambda + \xi$, $U_\pm$ the diagonalizing matrix of ${\cal M}^2 \pm F \lambda + \xi$ and there is an additional contribution to the sfermion scale
\be
\Delta \Lambda_S^2 = 4 {\rm Tr} \left[d \ \xi \ \log \frac{{\cal M}^2}{\mu^2}\right]\label{correccion}
\ee
This is a pure D-term contribution and is not definite positive, so it can largely alter the relation amongst sparticles. In \cite{Buican:2008ws} it was shown that this type of D-term contributions can enable the possibility of covering  the general GMSB spectrum.

In the limit in which the eigenvalues $(F\lambda \pm \xi)_k$ are small compared to the eigenvalues $(m^0_k)^2$, from equations (\ref{LambdaG}), (\ref{LambdaS}) and (\ref{correccion}) we obtain
\be
\Lambda_G = \frac{1}{2} \sum_{k = 1}^N \sum_\pm 2 d_{kk} \frac{(F\lambda \pm \xi)_k}{m_k^0}\ , \ \ \ \ \ \ \
\Lambda_S^2 = \frac{1}{2} \sum_{k = 1}^N \sum_\pm 2 d_{kk} \left[ \frac{(F\lambda \pm \xi)_k^2}{(m_k^0)^2} + 2 \xi_{kk} \log{(m^0_k)^2} \right]
\ee
As before, we can argue that this is a good approximation except when $(F\lambda \pm \xi)_k \approx (m_k^0)^2$.
Notice that if $F \lambda$ and $\xi$ are both simultaneously diagonalizable, these expressions become to lowest order
\be
\Lambda_G =  \sum_{k = 1}^N  2 d_{kk} \frac{F\lambda_k}{m_k^0}\ , \ \ \ \ \ \ \
\Lambda_S^2 = \sum_{k = 1}^N  2 d_{kk} \left(\frac{F^2\lambda_k^2}{(m_k^0)^2} + 2 \xi_{k} \log{(m^0_k)^2}\right) \equiv \Lambda_{S F}^2 + \Lambda_{S \xi}^2\label{lowest}
\ee
where $\xi_k$ are the eigenvalues of $\xi$.

Let us illustrate from equation (\ref{lowest}) for a $U(1)$ gauge group that these models cover the general GMSB spectrum. In this case, the parameter space of general GMSB is two dimensional and parameterized by $\Lambda_G$ and $\Lambda_S^2$. Following \cite{Buican:2008ws}, one can define the quotient $\kappa$
\be \kappa = \frac{\Lambda_S^2}{\Lambda_G^2}\ee
and show that when there are more than one messenger, $\kappa$ spans $\mathbb{R}$.
In fact, specializing equation (\ref{bounds2}) for a $U(1)$ gauge group, it can be seen that when there is only one messenger $\kappa$ is bounded $\kappa \in (0.37, 1)$. When there is more than one messenger it can be seen from (\ref{lowest}) that arbitrarily large values of $|\kappa|$ can be obtained setting $\xi = 0$. This is achieved by  adjusting the values and signs of the $\lambda_k$'s in such a way that $\Lambda_G$ is small, while $\Lambda_S$ is finite. On the other hand, with nonzero $\xi$ we can make $|\kappa|$ arbitrarily small, by taking $\Lambda^2_{S \xi} \approx - \Lambda^2_{S F}$.

 This is a strong motivation for analyzing these general models. In fact, in \cite{Buican:2008ws} it was shown that the extension from this $U(1)$ toy model to the physically relevant case of $G_{SSM}$ is possible. This requires messengers in at least three different irreps, and either D-breaking or arbitrary amount of SUSY breaking. The generalized messengers analyzed in this paper include such features, and should then be considered as a starting point to analyze complete and viable specific models of general GMSB. Implementing general GMSB with D-term breaking was also addressed in \cite{Carpenter:2008rj}.

\section{Summary and discussion} \label{Conclu}

We consider in this work a messenger sector with a generic mass term defined in (1)
which includes all couplings consistent with renormalizability and gauge invariance and
encompasses many of those found in the literature. The gauge group can be general, and the messengers can lie in different representations. To avoid dangerous negative contributions to the sfermion masses, we constraint the superpotential by imposing a messenger parity.
Through a recently introduced formalism \cite{Meade:2008wd}, we derive expressions for the soft masses of gauginos (\ref{LambdaG}), and of sleptons and squarks (\ref{masafermiones})-(\ref{LambdaS}). The formalism needs as unique  input from the hidden sector some correlators for the components of the gauge current superfield. The detailed computation is presented in the Appendix.

We then consider two different limits. One consists in taking the splitting of bossonic messenger masses to be small compared with the typical scale of the fermionic messenger masses (we refer to this as the ``small SUSY breaking limit''), in which case we recover the results of \cite{Dvali:1996cu}, \cite{Dimopoulos:1996ig}, \cite{Giudice:1997ni}. The other one is the limit in which ``effects due to multiple messenger scales'' can be ignored, in which case we assume that the eigenvalues of the messenger mass matrix are almost degenerate $m_n^0 \sim m_k^0$, $\forall k,n$. In this case, we obtain simple expressions for the soft scales that are manageable and can be predictive (\ref{ExpansionFGauginos}), (\ref{ExpansionFSfermions}), and generalize those of \cite{Cheung:2007es}.

After constraining the messenger sector with a global symmetry under which the spurion field is charged, we recover the mass expressions of Extraordinary Gauge Mediation previously obtained in \cite{Cheung:2007es} through a generalization of the Wave-function Renormalization Technique \cite{Giudice:1997ni}. Our masses generalize those, being valid beyond the small SUSY breaking limit, and we show that the deviation from this limit is analogous to that in the Minimal Gauge Mediation case \cite{Martin:1996zb}. Namely, the physics does not seem to change much, except when the SUSY breaking parameters $F \lambda_k / (m_k^0)^2$ are close to $1$. Some bounds are obtained for the masses (\ref{bounds})-(\ref{bounds2}), generalizing those in \cite{Martin:1996zb}.

Finally, we derive a correction to the soft masses (\ref{correccion}) induced by D-term effects. These effects are encoded in the matrix $\xi$ of Section \ref{Dterm}, which we have constraint to have vanishing trace ${\rm Tr} [d\ \xi] = 0$ since this translates into vanishing messenger supertrace, thus avoiding sensitivity to UV physics. These additional effects promote the messenger sector to be the most general one that can be constructed after F and D-breaking. We then provide complete formulas for the soft masses, after integration of the general messenger sector. We specialize the expressions in the limit of small SUSY breaking, and reproduce the results of \cite{Buican:2008ws}, showing for a toy example that it fully spans the general GMSB spectrum \cite{Meade:2008wd}.

There are many routes for further research. One would be to analyze the effects of multi-messenger scales and verify if they are actually ignorable or not; it would also be interesting to look for mechanisms to suppress these effects. Analyzing dynamical mechanisms to enforce messenger parity and real parameters in the messenger sector would also be very interesting. Some work in this direction was done in \cite{Dine:2007dz}, \cite{Carpenter:2008wi}.

Perhaps the most interesting continuation of this work would be to explore models leading to messenger mass matrices like those of (\ref{matrizconxi}), and then analyze how much they span the general GMSB spectrum of the Supersymmetric Standard Model, trying to find those that cover it completely. A motivation for this are the results of \cite{Buican:2008ws}, where it was demonstrated that spanning the whole parameter space is possible and that there are multiple ways to do so.

 Let us finally mention that our investigation was focused on the case in which, in the limit of vanishing coupling constants, the observable and hidden sectors are completely decoupled. It would then be of interest to study direct couplings between messenger and matter fields. In particular, couplings between the Higgs field and messenger doublets or singlets can lead to solutions to the $\mu$ problem (see \cite{Komargodski:2008ax} and references therein).

%%%%%%%%%%%%%%%%%%%%%%%%%%%%%%%%%%%%%%%%%%%%%%%%%%%%%%%%%%%%%%%%%
\vspace{1 cm} \noindent\underline{Acknowledgements}: I would like to thank Fidel Schaposnik for helpful discussions and comments. I also thank Bryan Leung for pointing out confusion notation. This work was partially supported by UNLP, UBA, and CONICET.
\newpage
%%%%%%%%%%%%%%%%%%%%%%%%%%%%%%%%%%%%%%%%%%%%%%%%%%%%%%%%%%%%%%%%%

\newpage

\section*{Appendix A. Avoiding tachyonic sfermions}

As was pointed out in \cite{Dvali:1996cu}, one-loop contributions to the masses of squark and sleptons arise in models in which the supersymmetric mass matrix $\cal M$ cannot be made diagonal in a basis in which the SUSY breaking one, $F \lambda$, is hermitic. These contributions come from a one-loop contraction of the hypercharge messenger D-term $D = g (\ti \phi^\dag Y \ti \phi -  \phi Y  \phi^\dag)$ and are phenomenologically unacceptable since they render the sfermion masses tachyonic. These terms are not generated up to three loops if there is a basis in which ${\cal M}$ is diagonal and $F \lambda$ hermitic, a fact that is not automatic but must be enforced for example through the so-called messenger parity \cite{Dvali:1996cu}, \cite{Dimopoulos:1996ig}. Another possibility to avoid this problem is that the so-called ``GUT-singlet hypothesis'' holds \cite{Dimopoulos:1996ig}.

In the absence of the superpotential (\ref{superpotential}) the theory would have an $SU(N) \times
\widetilde{SU(N)}$ messenger invariance transforming as
\be \phi \to D^*\ \phi_D \ , \ \ \ \ \ \ti\phi \to \ti D\ \ti\phi_D
\ , \ \ \ \ \  D \times \ti D \in SU(N) \times
\widetilde{SU(N)}\ee
This symmetry is broken by (\ref{superpotential}), but we can use it to fix a basis for $\cal M$
\be {\cal M}_D = D^\dag {\cal M} \ti D\ee
The theory is also invariant under a messenger parity (this symmetry is broken by ordinary particles)
\be \phi_D \to U_D^*\ \ti \phi_D^* \ , \ \ \ \ \ \ti\phi_D \to \ti
U_D\ \phi_D^* \ , \ \ \ \ \ V \to - V \ , \ \ \ \ \ \forall U_D, \ti
U_D \in SU(N)\ee
provided that for some given $U_D$, $\ti U_D$, we have
\be {\cal M}_D^\dag = U_D^\dag {\cal M}_D \ti U_D\ , \ \ \ \ \ \ F
\lambda_D^\dag = U_D^\dag F \lambda_D \ti U_D\ee
Then, imposing a messenger parity defined by $U_D = \ti U_D = 1$, we
guarantee that there is a basis $D \times \ti D$ in which ${\cal M}_D$ and
$F \lambda_D$ are both hermitic, and in particular $D\times \ti D$ can be chosen so as to diagonalize ${\cal M}$. In this paper we impose this symmetry and when deriving the mass spectrum we find it convenient to stand in the basis in which $\cal M$ is diagonal and real and $F \lambda$ hermitic.

\section*{Appendix B. Soft spectrum from the general GMSB formalism}

In this appendix we use the general GMSB formalism introduced in \cite{Meade:2008wd} to compute the soft masses for gauginos and sfermions. The advantage of this formalism is that one can avoid computing all the Feynman diagrams, and simply compute correlators for the components of the gauge current superfield, since they encode all the information needed from the hidden sector.

We begin by defining the lagrangian for the messengers coupled to the vector fields
\be \delta {\cal L} = \int d^2\theta d^2\bar\theta \left(\phi^\dag_i
e^{2 g V^a T^a} \phi_i + \ti\phi^\dag_i e^{-2 g V^a T^a}
\ti\phi_i\right) + \left(\int d^2\theta\ W +
c.c.\right) \label{hiddensector} \ee
where the superpotential $W$ is given by (\ref{superpotential}). All our computations are done in a basis in which $\cal M$ is diagonal and $F \lambda$ hermitic. For simplicity in this Appendix we consider all messengers in the same representation of a single gauge group. The generalization to multiple representations and gauge groups is
straightforward. Now we define the  real gauge current superfield ${\cal J}$  (satisfying  $\bar D^2 {\cal J}^a = D^2 {\cal J}^a = 0$) as
\be {\cal J}^a = \left.\partial_{2 g V_a} \left(\phi^\dag_i e^{2 g V^a T^a}
\phi_i + \ti\phi^\dag_i e^{-2 g V^a T^a} \ti\phi_i\right)\right|_{g
= 0} = \phi^\dag_i T^a \phi_i - \ti\phi^\dag_i T^a \ti\phi_i
\label{supercurrentdef}\ee
In components it reads
\be {\cal J}^a = J^a + i \theta j^a - i \bar \theta  \bar j^a -
\theta\sigma^\mu\bar \theta j_\mu^a +
\frac{1}{2}\theta\theta\bar\theta\bar \sigma^\mu\partial_\mu j^a -
\frac{1}{2}\bar\theta\bar\theta\theta \sigma^\mu\partial_\mu \bar
j^a - \frac{1}{4}\theta\theta\bar\theta\bar\theta \Box J^a\ee
where
\bea
J^a &=& \phi_i^\dag T^a \phi_i - \ti \phi^\dag_i T^a \ti \phi_i \nn\\
j^a &=& - i \sqrt{2} \left(\phi_i^\dag T^a \psi_i - \ti \phi^\dag_i T^a \ti \psi_i\right) \nn\\
\bar j^a &=& i \sqrt{2} \left(\bar \psi_i T^a \phi_i - \bar{\ti \psi}_i T^a \ti \phi_i\right) \nn\\
j_\mu^a &=& \left(\psi_i \sigma_\mu T^a \bar \psi_i - \ti \psi_i
\sigma_\mu T^a \bar{\ti\psi}_i\right) - i \left( \phi_i^\dag T^a
\partial_\mu \phi_i - \partial_\mu\phi^\dag_i T^a \phi_i -
\ti \phi_i^\dag T^a\partial_\mu \ti \phi_i + \partial_\mu\ti
\phi^\dag_i T^a \ti \phi_i\right) \label{supercurrent}\eea
We denote the chiral fields and their lowest component with the same letter $\phi$. The two-point correlation functions for the components of $\cal J$ are given by
\bea \langle J^a (0)\rangle &=& 0 \\
\langle J^a(x) J^b(0)\rangle &=& 2\ d \delta^{ab}\ {\rm Tr}
\left[ \Delta^+(x) \Delta^-(x)\right] \label{JaJb}\\
\langle j^a_\alpha(x)j^b_{\beta}(0)\rangle &=& - 2\ d \delta^{ab}\ {\rm Tr} \left[ (\Delta^+(x) - \Delta^-(x)) \epsilon_{\alpha
\beta} \ti \Delta^0(x)\right] \\
\langle j^a_\alpha(x) \bar j^b_{\dot \alpha}(0)\rangle &=& - 2 i\
d \delta^{ab}\ {\rm Tr} \left[ (\Delta^+(x) + \Delta^-(x))
\sigma^\mu_{\alpha \dot\alpha} \partial_\mu \Delta^0(x)\right] \\
\langle j^a_\mu(x) j^b_\nu(0)\rangle &=& 2\ d \delta^{ab}\
{\rm Tr} \left[ \sum_{\pm } \left(\partial_\mu \Delta^\pm(x)
\partial_\nu\Delta^\pm(x) - \Delta^\pm(x)\partial_\mu\partial_\nu
\Delta^\pm(x)\right) \right.  \label{CorrelatorMuNu}\\  && \ \ \ \ \
\ \ \ \ \ \ \ \ \ \ \ \ \ \  \ \ \ \ \left. +\ 2 \eta_{\mu\nu}
\left(\partial^\rho \Delta^0(x)
\partial_\rho \Delta^0(x) - \ti \Delta^0(x) \ti \Delta^0(x) \right) - 4 \partial_\mu \Delta^0(x)
\partial_\nu\Delta^0(x) \vphantom{\sum_{\pm }}\right]  \nn \eea
where $d \delta^{ab} = {\rm Tr} [T^a T^b]$ is the Dynkin index of the messengers, which is normalized to $1/2$ for
$SU(N_c)$ fundamentals. The trace runs in messenger space,  and we have defined the (rotated) propagators
\bea \Delta^{0}_{lm}(x) &=& \delta_{lk} \int \frac{d^4 p}{(2\pi)^4}
e^{i p x} \frac{i}{p^2 - (m^0_{k})^2} \ \delta_{km}\\
\ti\Delta^{0}_{lm}(x) &=& \delta_{lk} \int \frac{d^4 p}{(2\pi)^4}
e^{i p x} \frac{i m^0_{k}}{p^2 - (m^0_{k})^2} \ \delta_{km} \\
\Delta^\pm_{lm}(x) &=& (U_\pm)_{lk}\ \int \frac{d^4 p}{(2\pi)^4}
e^{i p x} \frac{i}{p^2 - (m^\pm_{k})^2} \ (U_\pm^\dag)_{km} \eea
in terms of $m^0_k$ and $(m^\pm_k)^2$, the eigenvalues of ${\cal M}$ and ${\cal M}^2 \pm F\lambda$ respectively, and $U_\pm$ the matrices that diagonalize ${\cal M}^2 \pm F\lambda$.
Now we implicitly define the functions $C_a(x^2 M^2)$ and $B_{1/2}(x^2 M^2)$ as follows
\bea
\langle J^a(x) J^b(0)\rangle &=&  \frac{1}{x^4} C_0(x^2 M^2)\ \delta^{ab} \label{C0}\\
\langle j^a_\alpha(x) \bar j^b_{\dot \alpha}(0)\rangle &=& - i \sigma^\mu_{\alpha \dot\alpha} \partial_\mu \left(\frac{1}{x^4} C_{1/2}(x^2 M^2)\right) \delta^{ab}\\
\langle j^a_\mu(x) j^b_\nu(0)\rangle &=& (\eta_{\mu\nu} \Box - \partial_\mu \partial_\nu) \left(\frac{1}{x^4} C_1(x^2 M^2)\right)  \delta^{ab}\\
\langle j^a_\alpha(x)j^b_{\beta}(0)\rangle &=&
\epsilon_{\alpha\beta} \frac{1}{x^5} B_{1/2}(x^2 M^2)\ \delta^{ab} \eea
Here $M$ is some characteristic mass scale of the theory (for instance $X$ in (\ref{XVEV})), which is introduced to express the arguments in terms of adimensional quantities. We also define the Fourier transformed functions $\ti C_a (p^2 / M^2; M / \Lambda)$ and $\ti B_{1/2}(p^2/M^2)$ as
\bea
\ti C_a (p^2/M^2; M /\Lambda) &=&  \int d^4 x e^{ipx} \frac{1}{x^4} C_a (x^2 M^2)\ =\   2 \pi^2 c \log(\Lambda/M)  + finite  \label{Ca}\\
M \ti B_{1/2} (p^2/M^2) &=& \int d^4 x e^{ipx} \frac{1}{x^5} B_{1/2}(x^2 M^2) \label{tiB}
\eea
Now, inserting the correlators (\ref{JaJb})-(\ref{CorrelatorMuNu}) in (\ref{C0})-(\ref{tiB}) one obtains (repeated indices are summed)
\bea
\ti C_0 &=& 2 d B_{kn} \int \frac{d^4q}{(2\pi)^4} \frac{1}{(q^2 + (m_k^+)^2)((p + q)^2 + (m_n^-)^2)} \label{tiC0}\\
\ti C_{1/2} &=& - \frac{2 d}{p^2} \sum_\pm  A^\pm_{kn} \int
\frac{d^4q}{(2\pi)^4} \frac{p\cdot q}{((p
+ q)^2 + (m_k^\pm)^2)(q^2 + (m_n^0)^2)} \label{tiC12}\\
\ti C_{1} &=& - \frac{2 d}{3 p^2} \int \frac{d^4q}{(2\pi)^4}\
\delta_{kn} \left[\sum_\pm \left(\frac{(p + q)\cdot (p + 2q)}{(q^2 +
(m_k^\pm)^2)((p + q)^2 + (m_k^\pm)^2)}- \frac{4}{q^2 +
(m_k^\pm)^2}\right) \right.\nn\\
&& \left. \ \ \ \ \ \ \ \ \ \ \ \ \ \ \ \ \ \ \ \ \ \ \ \ \ \ \ +
\frac{4 q\cdot(p + q) + 8 (m_k^0)^2}{(q^2 + (m_k^0)^2)((p
+ q)^2 + (m_k^0)^2)} \right] \label{tiC1}\\
M \ti B_{1/2} &=& 2 d \sum_\pm \mp A^\pm_{kn} \int
\frac{d^4q}{(2\pi)^4} \frac{m_n^0}{(q^2 + (m_k^\pm)^2)((p +
q)^2 + (m_n^0)^2)} \eea
where we have defined the messenger-rotating matrices\footnote{For a $U(1)$ gauge group, these correlators coincide with the results of \cite{Buican:2008ws} after the imposition of a messenger parity.}
\be B^\pm_{kn} = (U^\dag_\pm U_\mp)_{kn} (U_\mp^\dag U_\pm)_{nk} \ , \ \ \ \ \
A_{kn}^\pm = (U^\dag_\pm)_{kn} (U_\pm)_{nk} \ee
The last term in the first line of (\ref{tiC1}) represents a specific choice of contact terms,
which ar set such that the currents satisfy the conservation equations (ward identities) in momentum space \cite{Meade:2008wd}.

If the messenger supertrace vanishes (as it is the case here), all $\ti C_a$ agree up to ${\cal O}(1 / p^2)$, which is expected if supersymmetry is spontaneously broken. Notice that one obtains from (\ref{Ca}) that
the logarithmic divergent terms of $\ti C_a$ lead to $c = 2d N /
(2\pi)^4$, and from it one finds the shift in the beta function due
to the presence of the hidden sector, namely
\be
b_{high} - b_{low} = \Delta b = - (2 \pi)^4 c = -2d N
\ee
On the other hand, $\ti B_{1/2}$ is finite and receives no
contributions from the UV mass scale.

From the $\ti C_a$ functions we define the quantities $A_a$ as
\bea A_0 &=& - \int \frac{d^4 p}{(2\pi)^4} \frac{1}{p^2}\ \ti C_0
\ =\ - 2 d B^+_{kn} G_2(m_k^+, m_n^-)\nn\\
A_{1/2} &=& 4 \int \frac{d^4 p}{(2\pi)^4} \frac{1}{p^2}\ \ti
C_{1/2}\ =\ 4 d \sum_\pm A^\pm_{kn}\left[ G_1(0) (G_0(m_k^\pm) -
G_0(m_n^0))+ G_2 (m_k^\pm, m_n^0) \right.\nn\\&& \ \ \ \ \ \ \ \ \ \
\ \ \ \  \ \ \ \ \ \ \ \ \ \ \ \ \ \ \ \ \ \ \ \ \ \ \ \
\ \ \ \ \ \ \left.  + ((m_k^\pm)^2 - (m_n^0)^2)G_3(m_k^\pm, m_n^0)\right]\nn\\
A_1 &=&  - 3 \int \frac{d^4 p}{(2\pi)^4} \frac{1}{p^2}\ \ti C_1 \ =
\ - d \delta_{kn} \sum_\pm \left[4 G_1(0)(G_0(m_k^\pm) - G_0 (m_k^0)) + G_2(m_k^\pm, m_k^\pm) + 2 G_2(m_k^0, m_k^0)\right. \nn\\
&&\ \ \ \ \ \ \ \ \ \ \ \ \ \ \ \ \ \ \ \ \ \ \ \ \ \ \  \ \ \ \ \ \
\ \ \ \ \ \ \ \ \ \ \left. +\ 4 (m_k^\pm)^2 G_3 (m_k^\pm, m_k^\pm) -
4 (m_k^0)^2 G_3(m_k^0, m_k^0)\right] \label{Aes}
 \eea
which we have expressed in terms of the following integrals
\bea G_0(m) &=& \int \frac{d^4 p}{(2\pi)^4} \frac{1}{p^2 + m^2} \label{G0}\\G_1(m) &=& \int \frac{d^4 p}{(2\pi)^4} \frac{1}{(p^2 + m^2)^2}\\
G_2(m_1, m_2) &=& \int \frac{d^4 p}{(2\pi)^4} \frac{1}{p^2} \int
\frac{d^4 q}{(2\pi)^4}  \frac{1}{(q^2 + m_1^2)((p + q)^2 +
m_2^2)}\\
G_3(m_1, m_2) &=& \int \frac{d^4 p}{(2\pi)^4} \frac{1}{p^4} \int
\frac{d^4 q}{(2\pi)^4}  \frac{1}{(q^2 + m_1^2)((p + q)^2 + m_2^2)}\label{G3}\\
G_4(m_1, m_2) &=& \int \frac{d^4 p}{(2\pi)^4} \frac{1}{p^2} \int
\frac{d^4 q}{(2\pi)^4}  \frac{1}{((q + p)^2 + m_1^2)(q^2 + m_2^2)^2} \label{G4}
\eea

Now we have all the necessary ingredients to compute the soft terms.
The gaugino mass-matrix is diagonal, and each entry reads
\be M_{\ti g} = g^2 M \ti B_{1/2} (0) = \frac{\alpha}{4 \pi}
\Lambda_G \ee
where
\be \Lambda_G = 2 d\ m^0_n \sum_\pm \pm A_{kn}^\pm \frac{(m_k^\pm)^2 \log ((m_k^\pm)^2/(m_n^0)^2)}{(m_k^\pm)^2 - (m_n^0)^2} \ee
This concludes our derivation of the gaugino masses.
The sfermion masses are obtained through the relation
\be m_{\ti f}^2 = g^4\ C_{\ti f}\ (A_0 + A_{1/2} + A_1) \label{masafermionconAes}\ee
Each $A_a$ in (\ref{Aes}) is expressed in terms of the divergent integrals (\ref{G0})-(\ref{G3}). Notice from (\ref{Aes}) that since all $\ti C_a$ agree up to ${\cal O}(1 / p^2)$, the sfermion masses (\ref{masafermionconAes}) are UV finite, even though the individual terms contributing to it are not.
Following \cite{Martin:1996zb}, we use the Dimensional Regularization (DR) scheme in which the dimension is $n = 4 - 2\epsilon$, and  an infrared regulator $m_\epsilon$ is introduced. All the terms proportional to $G_0$ and $G_1$ in (\ref{Aes}) vanish due to the messenger supertrace formula, and the identity
\be (-1 + 2\epsilon)\ G_2(m_1, m_2) = m_1^2\ G_4(m_2,m_1) + m_2^2\
G_4(m_1,m_2) \ee
can be used to express the quantities $A_a$ exclusively in terms of the adimensional integrals $G_3$ and $G_4$, which after DR read
\bea
G_3(m_1, m_2) &=& \frac{\Gamma(1 + 2\epsilon)}{2 (4\pi^n)}\left[\frac{1}{\epsilon^2} + \frac{1}{\epsilon} (1 - 2 \log m_\epsilon^2) + 1 - \frac{\pi^2}{6} - F_2(m_1^2, m_2^2) - 2 F_3(m_1^2, m_2^2)\right. \nn\\
&& \ \ \ \ \ \ \ \ \ \ \ \ \ \ \ \left.   + 2 [F_1(m_1^2, m_2^2)-1]
\log m_\epsilon^2 + \log^2 m_\epsilon^2
\vphantom{\frac{1}{1}}\right] + {\cal O} (\epsilon)\\
G_4(m_1, m_2) &=& \frac{\Gamma(1 + 2\epsilon)}{2 (4\pi^n)}\left[\frac{1}{\epsilon^2} + \frac{1}{\epsilon} (1 - 2 \log m_2^2) + 1 - \frac{\pi^2}{6} - 2 \log m_2^2 + \log ^2 m_2^2\right. \nn\\
&& \ \ \ \ \ \ \ \ \ \ \ \ \ \ \ \left. -\log^2 m_1^2 + 2 \log m_1^2
\log m_2^2 - 2 Li_2\left(1 - \frac{m_2^2}{m_1^2}\right)\right] +
{\cal O} (\epsilon) \eea
Here we have defined the functions
\bea
F_1 (m_1^2, m_2^2) &=& (m_1^2 \log m_1^2 - m_2^2 \log m_2^2) / (m_1^2 - m_2^2)\nn\\
F_2 (m_1^2, m_2^2) &=& (m_1^2 \log^2 m_1^2 - m_2^2 \log^2 m_2^2) / (m_1^2 - m_2^2)\nn\\
F_3 (m_1^2, m_2^2) &=& (m_1^2 Li_2(1 - m_2^2 / m_1^2) - m_2^2 Li_2(1 -  m_1^2 / m_2^2)) / (m_1^2 - m_2^2)
\eea
when $m_1^2 \neq m_2^2$, and otherwise
\bea
F_1 (m^2, m^2) &=& 1 + \log m^2 \nn\\
F_2 (m^2, m^2) &=& 2 \log m^2 + \log^2 m^2 \nn\\
F_3 (m^2, m^2) &=& 2
\eea
$Li_2$ being the dilogarithm function.
Finally, from (\ref{masafermionconAes}) we obtain after some algebra and repeated use of the identity
\be
Li_2(-x) = - Li_2\left(\frac{x}{1 + x}\right) - \frac{1}{2} \log^2(1 + x) \label{DiLogProp}
\ee
the desired formula for the sfermion masses
\be m_{\ti f}^2 = 2 \left(\frac{\alpha}{4\pi}\right)^2 \ C_{\ti f}\ \Lambda_S^2 \label{masafermion}\ee
\be \Lambda_S^2 = 2 d  \sum_\pm (m_k^\pm)^2
\left[\ A^\pm_{kn} \log \frac{(m_k^\pm)^2}{(m_n^0)^2} - 2\ A^\pm_{kn}\ Li_2
\left(1 - \frac{(m^0_n)^2}{(m_k^\pm)^2}\right) +\ \frac{1}{2}\ B^\pm_{kn}\ Li_2 \left(1 - \frac{(m_n^\mp)^2}{(m_k^\pm)^2}\right)
\right] \label{Sfermionmass} \ee

%%%%%%%%%%%%%%%%%%%%%%%%%%%%%%%%%%%%%%%%%%%%%%%%%%%%%%%%%%%%%%%%%
\end{document}